\documentclass{ws-ijmpa}

\newcommand{\be}{\begin{equation}}
\newcommand{\ee}{\end{equation}}
\newcommand{\bea}{\begin{eqnarray}}
\newcommand{\eea}{\end{eqnarray}}
\newcommand{\bean}{\begin{eqnarray*}}
\newcommand{\eean}{\end{eqnarray*}}

\newcommand{\gapproxeq}{\lower
.7ex\hbox{$\;\stackrel{\textstyle >}{\sim}\;$}}
\newcommand{\lapproxeq}{\lower
.7ex\hbox{$\;\stackrel{\textstyle <}{\sim}\;$}}

\begin{document}

\markboth{Q. Zhao} {Charmonium hadronic decays and OZI rule
violation effects}

%
\catchline{}{}{}{}{}
%

\title{Charmonium hadronic decays and the OZI rule violation effects}

\author{Qiang Zhao}

\address{Theory Division, Institute of High Energy Physics, \\ Chinese Academy of Sciences, Beijing 100049 P.R. China \\
and Department of Physics, University of Surrey, Guildford, GU2
7XH, United Kingdom}

\maketitle

\begin{history}
\received{Day Month Year} \revised{Day Month Year}
\end{history}

\begin{abstract}
We discuss the scalar meson mixing scenario and present an OZI
rule violation mechanism for understanding the scalar productions
in charmoium hadronic decays. We stress that the OZI violation
could play a key role in disentangling the structure of the
scalars: $f_0(1370)$, $f_0(1500)$ and $f_0(1710)$.

\end{abstract}

\ccode{PACS numbers: 12.39.Mk, 13.30.Eg}

\section{About the scalars}

The successful quark model classification for the pseudoscalar and
vector nonet provides a natural reference for the expectation of
masses of the scalar $q\bar{q}$ nonet. Due to the orbital angular
momentum excitation which combines the symmetric spin to form
$0^{++}$, the quark model $q\bar{q}$ nonet is expected to have
masses heavier than their pseudoscalar and vector counterparts,
i.e. above 1 GeV. In this sense, the existence of scalars below 1
GeV, i.e. $\sigma(600)$ and $f_0(980)$, is already a sign showing
the non-trivial property of QCD. So far, more and more evidence
suggests that $\sigma(600)$, $f_0(980)$, $a_0(980)$ and
$\kappa(800)$ could be Jaffe's four-quark nonet~\cite{jaffe}.
Between 1$\sim$2 GeV, three scalars, $f_0(1370)$, $f_0(1500)$ and
$f_0(1710)$, were observed at DM2, MarkIII, WA102, Crystal
Barrel~\cite{bugg}. They were confirmed at BESII in $J/\psi$
radiative and hadronic decays. Nonetheless, BES also reported
signals of $f_0(1790)$ in $J/\psi\to \phi\pi\pi$ and $f_0(1810)$
in $J/\psi\to\gamma f_0(1810)\to \gamma\omega\phi$, which are
different in decay modes. The over-crowded isospin-0 scalars thus
bring difficulties to the interpretation and classification of the
scalar spectrum and raise questions on the non-perturbative QCD
dynamics.

In this proceeding, we will focus on $f_0(1370)$, $f_0(1500)$ and
$f_0(1710)$, for which more experimental information is available.
An essential consideration is that the Lattice QCD predicts the
lightest glueball to be $0^{++}$ with mass between 1.5$\sim$ 1.7
GeV~\cite{mp,ukqcd,chen}. This makes the scalars in this energy
region an natural candidate for the scalar glueball. However, due
to possibly large glueball-$q\bar{q}$ mixings, decisive evidence
is still unavailable. Recent data expose unexpected phenomena
which could be a chance for us to gain deeper insights into this
long-standing issue.

\begin{figure}
\begin{center}
\epsfig{file=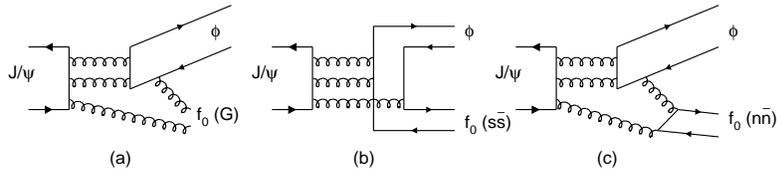, width=12cm,height=5.cm} \caption{Production
of $f_0^i$ in $J/\psi\to\phi f_0^i$. } \protect\label{fig-1}
\end{center}
\end{figure}

The systematic information from BES about these three states
turned out to be out of expectations: i) $f_0(1370)$ which was
seen in its strong decays into $\pi\pi$ and $4\pi$, and hence
deducted to have a large $n\bar{n}\equiv
(u\bar{u}+d\bar{d})/\sqrt{2}$, is found to be produced preferably
via recoiling $\phi$ instead of $\omega$ in $J/\psi\to V f_0$,
where $V=\omega, \ \phi$; ii) $f_0(1710)$ which couples to
$K\bar{K}$ strongly, is found produced preferably via recoiling
$\omega$ instead of $\phi$. iii) $f_0(1500)$ is not directly seen
in the invariant mass spectrum though it is needed in the partial
wave analysis. The ``scalar puzzle" arises from the OZI-rule
expectation of the production of the scalars: it favors to occur
via the singly OZI disconnected processes (SOZI) as illustrated by
Fig.~\ref{fig-1}(a), while the doubly OZI disconnected process
(DOZI, see Fig.~\ref{fig-1}(b)) should be strongly suppressed. In
this sense, the puzzle also raises the question about the scalar
production mechanism in $J/\psi$ hadronic decays. As follows, we
will clarify the correlations between the scalar mixings and their
production mechanisms. We will show that large OZI violations are
expected in association with the scalars, and can be tested in
experiment.

\section{OZI-rule violation in charmonium hadronic decays}

In Refs.~\cite{close-amsler,close-kirk}, a favor mixing scheme was
proposed to accommodate those three scalars in a flavor-singlet
basis of glueball and $q\bar{q}$. By fitting the data for $f_0\to
PP$, where $P$ stands for pseudoscalar mesons, a mixing pattern of
the glueball and $q\bar{q}$ components is highlighted, and allows
qualitative expectations of the scalar decays into other channels
such as radiative and two-photon decays. In
Ref.~\cite{close-zhao}, an improved calculation for $f_0\to PP$ is
presented in association with the new data from $J/\psi\to V f_0$
transitions~\cite{bes}. It is found that $f_0(1370)$ has the
largest $(u\bar{u}+d\bar{d})/\sqrt{2}$ component, while
$f_0(1710)$ has the largest $s\bar{s}$. The glueball component in
$f_0(1500)$ turns out to be the largest though it is also sizeable
in $f_0(1710)$.

To be consistent with the $G$-$q\bar{q}$ mixing scheme, an
essential question arising from $J/\psi\to V f_0$, is on the role
played by the DOZI processes. As shown in Ref.~\cite{close-zhao},
a sizeable contribution from the OZI violation can explain the
features observed in the production of those three scalars.
Although the numerical results are qualitative, the interfering
pattern among the SOZI and DOZI processes as a result of the
$G$-$q\bar{q}$ mixings is rather stable. This naturally leads to
the following concerns: i) What is the role played by the
non-perturbative DOZI processes in $J/\psi\to V f_0$? ii) What is
the correlation between the $G$-$q\bar{q}$ mixings and the DOZI
processes in the scalar production?

Interestingly, the different considerations of the role played by
the DOZI processes in $J/\psi$ hadronic decays seem to be one of
the major divergences in different phenomenological approaches. To
be specific, negligibly small DOZI contributions imply the
dominance of pQCD transitions, and also small $G$-$q\bar{q}$
mixings. Based on this, a criteria pointed out by
Carlson~\cite{carlson-81} and recently developed by
Chanowitz~\cite{chanowitz-2005} is the chiral suppression
mechanism for $J=0$ glueballs. Due to the fact that in pQCD the
amplitude is proportional to the current quark mass in the final
states, the $J=0$ glueballs will have larger couplings into e.g.
$K\bar{K}$ rather than $\pi\pi$. For $J\neq 0$, the decay
amplitude is flavor symmetric. In association with the lattice QCD
calculations, $f_0(1710)$ is thus proposed to be a glueball
candidate. However, as pointed by Chao {\it et al.}, the chiral
suppression does not materialize itself in the hadronization
process. Thus, the observation of relatively larger b.r. to
$K\bar{K}$ for a candidate does not necessarily lead to its being
a glueball~\cite{chao-he-ma,zhang-jin}.

In contrast, the study of Isgur and Geiger by calculating hadronic
loop contributions to meson propagators suggests that large OZI
violations occur in the $0^{++}$ sector while systematic
cancellations are found in other nonets~\cite{isgur-geiger}. A
general argument for the large OZI violations in $0^{++}$ is
provided by Lipkin and Zou~\cite{lipkin-zou}. QCD sum rule
calculations also support large OZI-rule violations in the
scalars~\cite{kisslinger,faessler,narison}. Unquenched lattice QCD
calculations of the glueball spectrum should be crucial for
clarifying the above two major scenarios.

In $J/\psi$ hadronic decays, some experimental evidence seems
available for large OZI violation effects. The reaction,
$J/\psi\to V f_0$, can occur via intermediate meson rescatterings,
i.e. $J/\psi\to K^*\bar{K}+c.c. \to V f_0$ and/or $J/\psi\to
\rho\pi\to V f_0$. Since $J/\psi\to K^*\bar{K}+c.c.$ and
$J/\psi\to \rho\pi$ are two of the largest decay channels of
$J/\psi$, the intermediate meson rescatterings as the dominant
contributions to the DOZI processes may not be small. Theoretical
study of such processes should be able to clarify the role played
by the DOZI processes in $J/\psi\to V f_0$ and provide insights
into the scalar structures.

\section{Intermediate meson rescatterings}

\begin{figure}
\begin{center}
\epsfig{file=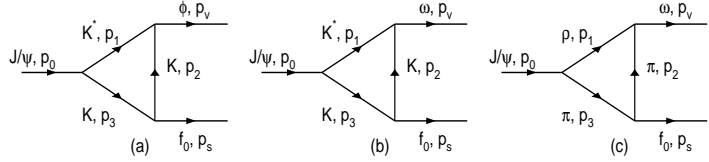, width=12cm,height=5.cm} \caption{The OZI
violations from intermediate $K^*K$ and $\rho\pi$ meson exchanges
as dominant contributions to the doubly disconnected processes. }
\protect\label{fig-2}
\end{center}
\end{figure}

Following the meson exchange mechanisms, we calculate the meson
loops in Fig.~\ref{fig-2} as a leading contribution to the doubly
disconnected process. The transition ampliude can be written as
\be
M_{fi}=-i\int\frac{d^4 p_2}{(2\pi)^4}T_v^\beta
\left(g^\lambda_\beta-\frac{p_{1\beta}p_1^\lambda}{p_1^2}\right)
T_{0\lambda} T_s\frac{F(p_2)}{a_1 a_2 a_3} \delta^4(p_0-p_v-P_s) \
,
\ee
where the vertex functions are
\bea
\label{vertex-func} T_v^\beta &= &
i\frac{g_v}{M_v}\epsilon^{\mu\nu\alpha\beta}
p_{v\mu}\epsilon_{f\nu}p_{2\alpha}, \nonumber\\
T_\lambda &=& i\frac{g_0}{M_0}
\epsilon_{\lambda\sigma\tau\delta}p_0^\sigma\epsilon_i^\tau p_3^\sigma , \nonumber\\
T_0 & =& ig_s M_s \ ,
\eea
and $a_1=p_1^2-m_1^2+i\epsilon$, $a_2=p_2^2-m_2^2 +i\epsilon$, and
$a_3 = p_3^2-m_3^2 +i\epsilon$ are the denominators of the meson
propagators. The coupling constants $g_v$, $g_s$ and $g_0$ can be
determined independently in meson decays. For the $J/\psi$
hadronic decays, large branching ratios for $J/\psi\to K^*\bar{K}
+ c.c.$ and $ \rho\pi +c.c.$ imply that the meson loop transitions
may have significant contributions to those decays where the final
state mesons have also large couplings to the exchanged mesons.
This is indeed the case for $J/\psi\to V f_0(1710)$, where the
couplings for $\phi K^*K$, $\omega K^*K$, and $f_0(1710)K\bar{K}$
are significantly large. Therefore, this simple argument will
allow us to consider only the dominant meson exchange loops in the
calculations.

In the above equation, the coupling $g_0$ can be determined by the
decays of $J/\psi \to K^*\bar{K}+c.c.$ [Fig.~\ref{fig-2}(a) and
(b)], or $J/\psi\to \rho\pi+c.c.$ [Fig.~\ref{fig-2}(c)], of which
large branching ratios are observed in experiment, e.g.
\be
g_0^2=\frac{12\pi M_0^2}{|{\bf p}_1|^3} \Gamma^{exp}_{J/\psi\to
V\bar{P}} \ ,
\ee
where $\Gamma^{exp}_{J/\psi\to K^*\bar{K}+c.c.}= (9.2\pm
0.8)\times 10^{-3}$ and $\Gamma^{exp}_{J/\psi\to
\rho\pi+c.c.}=(1.27\pm 0.09)\% $ are from the estimate of Particle
Data Group~\cite{pdg2004}.

For  $f_0\to K\bar{K}$ in the $f_0$ c.m. system, the coupling
constant can be derived via $ g_s^2={8\pi}\Gamma^{exp}_{f_0\to
K\bar{K}}/{|{\bf k}|}$, which is consistent with the studies of
$f_0\to PP$ in the determination of the mixing matrix
elements~\cite{close-zhao}; $|{\bf k}|$ is the magnitude of the
three momenta carried by the final-state kaon (anti-kaon). With
the estimate of $b.r._{f_0\to K\bar{K}}=0.60$~\cite{close-zhao},
and $\Gamma_T(1710)=140$ MeV~\cite{pdg2004}, we have
$\Gamma^{exp}_{f_0\to K\bar{K}}=84$ MeV; the coupling $g_s$ can
then be determined. We determine the $VVP$ couplings, i.e.,
$g_{\phi K^* K}$, $g_{\omega K^*K}$, and $g_{\omega \rho\pi}$, in
the SU(3)-flavor-symmetry limit with $g^2_{\omega
\rho^0\pi^0}\simeq 84$ is determined in vector meson dominance
(VMD) model in $\omega\to \pi^0 e^+ e^-$~\cite{pdg2004}.

\begin{table}[ph]
\tbl{The intermediate meson exchange contributions to the decay of
$J/\psi\to V f_0\to V PP$ with a dipole form factor. The data are
from BES.} {\begin{tabular}{c|c|c|c} \hline
$(\times 10^{-4})$ & $f_0(1710)$ & $f_0(1500)$ & $f_0(1370)$ \\[1ex]
\hline
$b.r.(J/\psi\to\phi f_0) $ & 1.73 &0.24 & 0.15 \\[1ex]
$b.r.(J/\psi\to\phi f_0\to \phi K\bar{K}) $ & 1.04 & 0.02 & 0.00
\\[1ex]
$b.r.$(exp) & $(2.0\pm 0.7)$ & $(0.8\pm 0.5)$ & $(0.3\pm 0.3)$
\\[1ex]\hline
$b.r.(J/\psi\to\omega f_0) $ & 1.43 & 0.19 & 0.11 \\[1ex]
$b.r.(J/\psi\to\omega f_0\to \omega K\bar{K}) $ & 0.86 & 0.02 &
0.00 \\[1ex]
$b.r.$(exp) & $(13.2\pm 2.6)$ & $\dots$ & $\dots$ \\[1ex]\hline
$b.r.(J/\psi\to\omega f_0) $ & 0.57 & 2.59 & 3.43 \\[1ex]
$b.r.(J/\psi\to\omega f_0\to \omega \pi\pi) $  & 0.04 & 0.90 &
0.69 \\[1ex]
$b.r.$(exp) & $\dots$ & $\dots$ & $\dots$\\[1ex] \hline
\end{tabular}
\label{tab-1}}
\end{table}

In Table~\ref{tab-1}, the intermediate meson rescattering
contributions to the branching ratios are listed. It shows some of
those transitions play an important in $J/\psi\to Vf_0$, and can
produce significant branching ratios compatible with the
experimental data. This can be regarded as an instructive hint
about the OZI violation effects in the $J/\psi$ hadronic decays.

The intermediate meson rescatterings can be compared with the
``tree" diagram for $J/\psi\to V\bar{P}+c.c.$, e.g. $J/\psi\to K^*
\bar{K}+c.c.$ and $J/\psi \to \rho\pi+c.c.$ Since the $J/\psi
V\bar{P}$ vertices are the same between the tree and loop
transitions, the ratio between the intermediate meson rescattering
loop and the tree process of $J/\psi\to V\bar{P}+c.c.$ will
highlight the OZI violations in this dynamical process and can be
related to the scalar flavor contents by measuring the following
fraction~\cite{zzm}:
\be
\label{ozi-ratio} R^{OZI}_i=\frac{\Gamma_{J/\psi\to\phi
f_0^i\to\phi K\bar{K}}} {\Gamma_{J/\psi\to\omega f_0^i\to\omega
K\bar{K}}} =\frac{| {\bf p}_{\phi i}|}{| {\bf p}_{\omega i} |}
\frac{[  x_i + y_i +\sqrt{2}r z_i ]^2}{2[ x_i  + r y_i + \sqrt{2}
z_i ]^2} \ ,
\ee
where $x_i$, $y_i$, and $z_i$ are the mixing angles of $G$,
$s\bar{s}$ and $(u\bar{u}+d\bar{d})/\sqrt{2}$ components for
scalar $i$; Parameter $r$ denotes the relative strength between
DOZI and SOZI transitions, which can be estimated by the ratio
between the intermediate meson rescattering loop and the tree
process. Since apart from the kinematic factors $| {\bf p}_{\phi
i}|$ and $| {\bf p}_{\omega i} |$, the only energy-dependent
factor in $R^{OZI}_i$ is $r$, one can thus measure $R^{OZI}$ in
both $J/\psi\to V f_0$ and $\Upsilon\to V f_0$, where $r$ is
expected to change from a sizeable value in $J/\psi$ decays to a
smaller one in $\Upsilon$ decays, to test the configuration of the
scalars~\cite{zzm}.

\section{Summary}

In brief, we discussed a possible way to determine the scalar
structures by clarifying the role played by the OZI-rule
violation. The latter was correlated with the glueball-$Q\bar{Q}$
mixings in $J/\psi\to V f_0^i$. Since the flavor wavefunctions for
$\omega$ and $\phi$ are almost ideally mixed, the decay channels
into $\omega$ and $\phi$ in association with the scalar mesons,
respectively, serve as a flavor filter for probing the $Q\bar{Q}$
contents of the scalars. This allows us to separate out the doubly
OZI disconnected processes,  of which the effects can be measured
by the branching ratio fractions between $\phi f_0^i$ and $\omega
f_0^i$, i.e. $R^{OZI}_i$. Since the energy evolution of
$R^{OZI}_i$ is mostly determined by the energy evolution of the
doubly disconnected processes relative to the singly disconnected
ones, the suppression of the doubly disconnected process at higher
energies, e.g. in $\Upsilon$ decays, will lead to dramatic changes
to $R^{OZI}_i$ with certain patterns. Observation of such a change
will provide direct information about the scalar meson structures.

\section*{Acknowledgement}

The author thanks M. Chanowitz, K.T. Chao, L. Kisslinger, and K.F.
Liu for many useful discussions. Collaborations with F.E. Close
and B.S. Zou on relevant works are acknowledged.

\end{document}